\documentclass[showpacs,showkeys,aps,prl,preprint,groupedaddress]{revtex4}

\usepackage[latin1]{inputenc}
\usepackage{amsmath} % need forsubequations
\usepackage{graphicx} % need for figures
\usepackage{hyperref} % use for hypertext links, including those to external documents and URLs

\begin{document}

% Use the \preprint command to place your local institutional report
% number in the upper righthand corner of the title page in preprint mode.
% Multiple \preprint commands are allowed.
% Use the 'preprintnumbers' class option to override journal defaults
% to display numbers if necessary
%\preprint{1}

%Title of paper
\title{Torque determination on DNA with magnetic tweezers}

% repeat the \author .. \affiliation  etc. as needed
% \email, \thanks, \homepage, \altaffiliation all apply to the current
% author. Explanatory text should go in the []'s, actual e-mail
% address or url should go in the {}'s for \email and \homepage.
% Please use the appropriate macro foreach each type of information

% \affiliation command applies to all authors since the last
% \affiliation command. The \affiliation command should follow the
% other information
% \affiliation can be followed by \email, \homepage, \thanks as well.
\author{Francesco Mosconi} \email[]{mosconi@lps.ens.fr} \author{Jean
  Fran\c ois Allemand} \author{David Bensimon} \author{Vincent
  Croquette}
%\thanks{}
%\altaffiliation{}
\affiliation{LPS, ENS, UMR 8550 CNRS, 24 rue Lhomond, 75231 Paris
  Cedex 05, France}\homepage[]{www.lps.ens.fr/recherche/biophysique-ADN}

%Collaboration name if desired (requires use of superscriptaddress
%option in \documentclass). \noaffiliation is required (may also be
%used with the \author command).
%\collaboration can be followed by \email, \homepage, \thanks as well.
%\collaboration{}
%\noaffiliation

\date{\today}

\begin{abstract}
We deduced the torque applied on a single stretched and twisted DNA  by integrating with respect to force the change in the molecule's extension as it is coiled.  While consistent with previous direct measurements of the torque at high forces ($F>1$pN) this method, which is simple and does not require a sophisticated set-up, allows for lower force estimates. We used this approach to deduce the effective torsional modulus of DNA, which decreases with force and to estimate the buckling torque of DNA as a function of force in various salt conditions.
\end{abstract}

% insert suggested PACS numbers in braces on next line
% insert suggested keywords - APS authors don't need to do this

%\maketitle must follow title, authors, abstract, \pacs, and \keywords
\maketitle

%\section{\label{sec:introduction}Introduction}
Most polymers are insensitive to torsion because their monomers are linked by  single covalent bonds around which they are free to rotate.
This property is lost when the polymer possesses no single covalent
bond about which to release the accumulated torsion. Such is the case of
a DNA molecule with no nicks (no break
in one of the strands), thanks to is double-helical structure. This particular feature has very important
biological implications. First, from a structural point of view,
twisted DNA  provides an efficient way to compact the molecule so that
it fits into the cell or nucleus. Second, a negatively twisted (underwound) DNA may
locally denature thus facilitating the accessibility of its bases to a variety of
proteins (RNA polymerases\cite{revyakin_promoter_2004}, regulation factors\cite{lia03}, etc.). On the other
hand, positively coiled DNA is more stable at high temperature (it
denatures less). Thus thermophilic bacteria that live close to the
boiling point of water have  enzymes  (reverse gyrases) that overwind
the molecule. Because the topology of DNA plays such an essential role
in the cell life, Nature has evolved a family of enzymes, generally known as topoisomerases\cite{stri00,koster_friction_2005,taneja_topoV_07}(the just mentioned reverse gyrase\cite{cozz80} is one of them) that control the torsion and entanglement of the molecules. Enzymes that translocate DNA (such as FtsK\cite{saleh_analysis_2005},EcoR124I\cite{seidel_motor_2008}, RSC\cite{lia_direct_2006}) can also apply a torque on the molecule as it is moved along. Thus understanding the behaviour of DNA under torsion and estimating the torque arising in a twisted molecule has important biological implications.

%The torsional state of DNA can be described using two geometrical variables : the {\it twist} ($Tw$) and the {\it Writhe} ($Wr$) and one topological variable, the {\it linking number ($Lk$)}. The twist $Tw$ is a geometrical variable that measures the number of time the two DNA strands wrap around each other. The natural twist $Tw_0$ of a DNA molecule equals the total number of basepairs (bps) $n_0$ divided by the average number of bps in a helical pitch $n_p=10.5 $bps: $ Tw_{0} =  n_0 / n_p $. The second geometrical variable, the writhe $Wr$, is related to the number of times the DNA's axis crosses itself, as when a cord is coiled to form plectonemes. Finally, the linking number $Lk$ is defined as the number of strand passages one has to execute in order to separate the two DNA strands. $L_k$ is a topological variable, i.e it is not affected by changing the geometry (bending, stretching, etc...) of a circular DNA, as long as no strand is cut. A mathematical theorem \cite{calu59,whit69} states that: $ Lk = Tw + Wr $. The degree of supercoiling of a DNA molecule $\sigma = n/Lk_{0}$ is defined as the number of extra turns $n$ (with respect to $Tw_{0}$) applied to the molecule divided by its native twist or linking number: $Lk_{0} = Tw_{0}$. In most bacteria the degree of DNA supercoiling is found to be about -0.06, i.e. the DNA is slightly unwound. The torque arising from this coiling of DNA is often used by the cell to control DNA processes such as transcription. It is thus important to be able to estimate that value.

Single molecule manipulation experiments\cite{stri00,koster_friction_2005,taneja_topoV_07, bryant_structural_2003, saleh_analysis_2005,seidel_motor_2008,lia_direct_2006} offer a means to stretch and twist DNA. In these experiments, a DNA molecule is anchored at multiple points (to impede
its  swiveling) to a surface at one end and to a bead used to apply a force and a torque at the other. In the case of magnetic traps, a
superparamagnetic bead is pulled by the field generated by small magnets and twisting is achieved by rotating the magnetic field\cite{strick_elasticity_1996,gosse_magnetic_2002}.
With this method the angular position of the bead is imposed and one does not control the applied torque. This set-up was nonetheless used to measure the twist-stretch coupling in a DNA molecule via the rotational drag of a small bead attached to the backbone and allowed to swivel to relax the accumulated torsion\cite{gore06}.
More recently, optical tweezers\cite{wang_stretching_1997} have been used to apply a constant torque on an anisotropic
transparent particle through the angular momentum transfer of a polarized laser beam\cite{deufel_nanofabricated_2007} that also traps the particle. The advantage of the optical tweezers set-up is that it allows for a direct measurement of the torque
applied on the trapped particle (and through it on the DNA). Its drawback is that it involves a rather sophisticated set-up which is difficult to use to explore the low force (below 1pN) regime that might be more relevant to biological processes.

\begin{figure}[!ht]
\includegraphics[width=9cm]{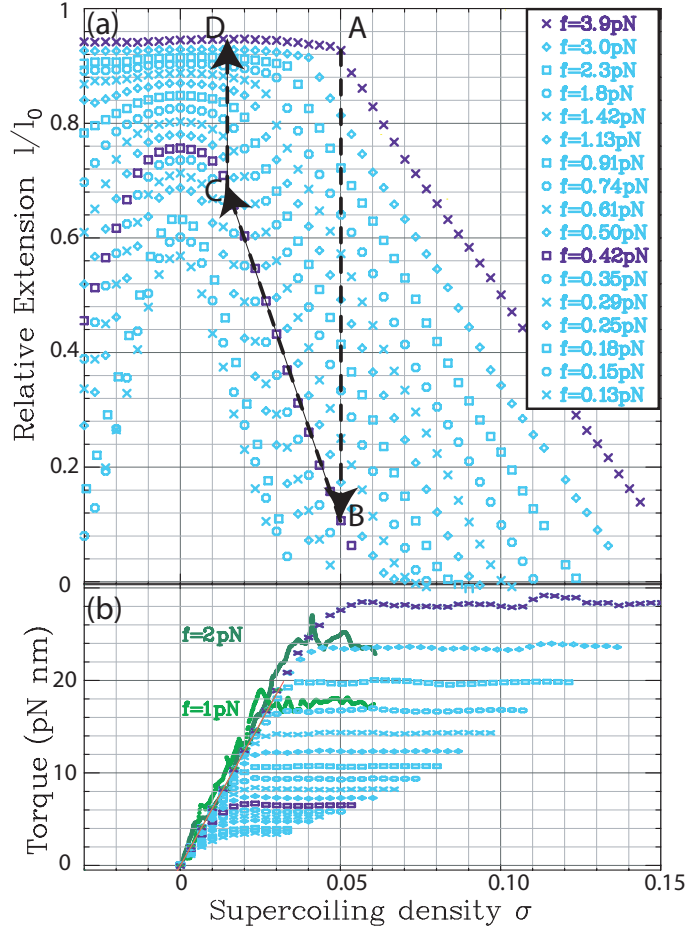}
\caption{\label{fig:hats} (a) Variation of the mean relative extension
  of a DNA molecule $l/l_0$ as a function of the degree of
  supercoiling $\sigma = n / Lk_0$  in 100mM NaCl. (b) Variation of
  the torque in DNA as a function of $\sigma$. These curves have been
  obtained by integrating as explained in the text Eq.\ref{eq:torque}
  from points A to B and from points C to D in (a) assuming that the
  torque in the plectonemic regime at a given force (i.e. at points B
  and C) is constant. The wiggly continuous lines in (b) are results
  from ref. \cite{forth_abrupt_2008}.}
\end{figure}

In this paper we describe a simple method to estimate the torque applied on DNA by measuring the change in extension of a stretched and coiled molecule as a function of force and number of turns. The magnetic trap system, briefly sketched above and employed extensively in previous experiments \cite{strick_elasticity_1996, stone_chirality_2003,bancaud_structural_2006,dawi06}, is used to apply a force $F$ on a magnetic bead tethered by a single DNA to a surface and also, by rotating the magnets, to twist the molecule by a known number of turns, $n$. The results of such experiments are qualitatively easy to understand on the basis of our daily experience with coiling tubes or cords. Consider twisting by $n$ turns a rubber tube of torsional modulus $\cal C$ (usually normalized by $k_B T$ in the DNA context: ${\cal C} = k_B T C$) held under an applied force
$F$.  Initially the torque $\Gamma$ will increase linearly with $n$:
$\Gamma = 2 \pi n  {\cal C}/{l} $, leaving the extension $l$ almost
unchanged. Past a certain number of turns $n_b$, the associated torque
$\Gamma_b$ becomes so large that it is energetically less costly for
the tube to bend rather than to increase its torsional energy: the tube buckles and loops to form a 3D structure called plectoneme, that absorbs torsion as writhe. Further twisting of the tube, while leaving the torque unchanged, results in formation of ever longer plectonemes. Very similar results are observed (see below and Fig.\ref{fig:hats}) when coiling a DNA molecule: while its extension $l$ varies little for small $n$ it decreases linearly past a certain threshold. As recently suggested
\cite{zhang_maxwell_2008}, from these observations one can deduce the torque applied on the molecule. Indeed the free energy $\mathcal F$ of a twisted and stretched molecule depends on the force $F$ and the rotational angle, $\theta = 2 \pi n$. The mean extension of the molecule at a given force is: $l = - \partial \mathcal F/\partial F|_\theta$ and the mean torque is: $\Gamma = \partial \mathcal F/\partial \theta|_F$. One thus readily derives an expression for the mean torque at a given force $\Gamma(F,\theta)$ from a measurement of the decrease in extension with increased coiling $\partial l / \partial \theta|_F$:

\begin{equation}
\label{eq:torque}
\Gamma(F,\theta)=\Gamma(F_0,\theta)-\int_{F_0}^F dF' \left(\frac{\partial l}{\partial \theta} \right)_{F'}
\end{equation}

Since the angular rotation is known and the force and change in extension with rotation are easily measurable, the determination of the torque difference is reduced to a problem of sampling these variables finely enough to estimate the above integral with sufficient precision. The integration constant $\Gamma(F_0,\theta)$ is set by the requirement that there is no torque on an untwisted molecule: $\Gamma(F,0) = 0$ (neglecting the small twist-stretch coupling modulus of DNA\cite{lion06,gore06}).

We measured the DNA extension (of total length $l_0 \approx 5.4 \mu$m) in various salt conditions and for different values
of force $F$ and degree of supercoiling $\sigma = n /Lk_{0}$ (the linking number $Lk_0 \approx 1500$ is the number of times the two
strands of the molecule wrap around each other), see Fig.\ref{fig:hats}. Such
measurements have been described before \cite{strick_elasticity_1996,strick_behavior_1998}. Briefly at low forces ($F$ less than about 0.4pN) the curves are symmetric. The extension is maximal at $\sigma=0$ and decreases non-linearly for small values of
$\sigma$ due to twist fluctuations
\cite{moroz_entropic_1998,bouchiat_estimating_1999}. Past the buckling threshold (at $\sigma =\sigma_s$), the molecule coils on
itself with a constant slope ($\partial l/\partial \sigma|_F$) to form
plectonemes or supercoils. As explained above, while below buckling
(i.e. when $\sigma < \sigma_s$) the torque increases with increased
rotation, it is constant in the plectonemic regime
\cite{marko_torque_2007} (i.e. when $\sigma > \sigma_s$). For larger
forces ($F>0.5$ pN) the curve becomes asymmetric as for negative
supercoilings DNA denatures before buckling at a critical torque:
$\Gamma_d \sim 9$ pN nm \cite{strick_micro-mechanical_1999,bryant_structural_2003}. For this reason we have computed the torque
only for positive degrees of supercoiling, although this estimate
should also be valid at low forces ($F<0.5$ pN) for negative supercoilings.

To compute the torque at various forces and degrees of supercoiling
using Eq.\ref{eq:torque}, we start from the highest stretching force
($F_{max} = 3.9$ pN in the series shown in fig.\ref{fig:hats}(a)) and
the highest torque state investigated, namely point A at the buckling
transition ($\sigma = \sigma_{s,max}$) in Fig.\ref{fig:hats}(a). The
value of torque in A (which served as our reference point) is
initially unknown but will be fixed by the requirement that
$\Gamma(F,0) = 0$. We then compute the torque at point D ( $\Gamma_D$
for which $\sigma_D < \sigma_{s,max}$), by first integrating
Eq.\ref{eq:torque} along path AB (a path of constant $\sigma =
\sigma_A$ but varying force, see Fig.\ref{fig:hats}(a)) from $F_{max}$
to $F_B$. Taking into account the fact that the torques at points C
and B are equal $\Gamma_C = \Gamma_B$), we then calculate the torque
difference along the path CD from force $F_C=F_B$ to force $F_D$ along
a path of constant $\sigma = \sigma_D$ and subtract it from the torque
difference along path AB. This procedure is of course valid only if
both points B and C are in the plectonemic regime. This in effect
restricts that procedure to values of $\sigma>0.02$, where we can identify correctly the buckling transition. To evaluate the
torque at smaller values, we notice that the values of $\Gamma$ as a
function of $\sigma$ at high forces grow linearly with $\sigma$. We
extrapolate the values of $\Gamma$ for one of these curves down to
$\sigma=0$ (requiring $\Gamma(F,0)=0$). It does not matter which curve
is used: the intercept with the ordinate at $\sigma=0$ varies by less
than 1 pN nm.  We then use these extrapolated values to infer from the
numerical integration of Eq.\ref{eq:torque} the values of the torque
at other forces and values of $\sigma <0.02$. The results are shown in
Fig.\ref{fig:hats}(b). Various methods to evaluate the derivative
$\frac{\partial l}{\partial \theta}|_F$ have been used and found to yield
very similar results (a Savitzky-Golay\cite{numerical} five points second order smoothing method
was usually preferred as it is less affected by noise due to discrete
sampling). The results obtained here are similar, see Fig.\ref{fig:hats}(b) to the results reported on a
different DNA molecule using optical tweezers as a means to measure
the torque (the slightly different values of the buckling torques
might be due to differences in the DNA sequences or to different ionic
conditions).

From the data in Fig.\ref{fig:hats}(a,b) we can deduce the effective
tension and torque on a bare plasmid (circular DNA) unwound by $6\%$ (as often found in Nature\cite{mccl90}). The tension is determined by the value of force at which the DNA's extension is zero at $\sigma = 0.06$, i.e. $F\sim 0.42$
pN\cite{char04} (the line passing through the points BC (at $F=0.42$ pN) crosses the abscissa at $\sigma=0.06$). The buckling torque at this force is obtained from Fig.\ref{fig:hats}(b): $\Gamma \sim 6$ pN nm $= 1.5 k_B T$. This value is close to the value where DNA denatures at negative supercoilings. This may explain the variable sensitivity of gene expression to sequence and degree of supercoiling\cite{revyakin_promoter_2004,drog94}.

\begin{figure}[!htb]
\includegraphics[width=9cm]{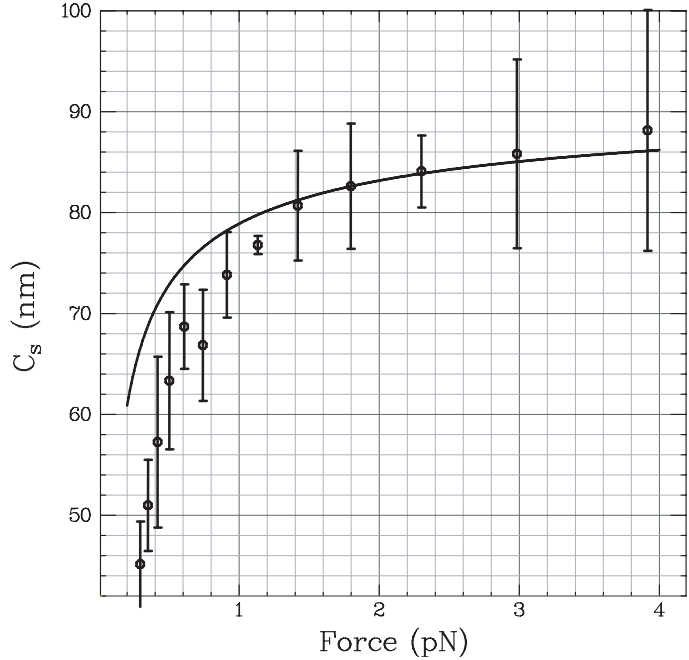}
\caption{\label{fig:ccomp} Variation with force of the effective torsional stiffness of DNA in 100mM NaCl. The continuous curve is a best fit of the high force data to the prediction of ref.\cite{moroz_torsional_1997}, with C = 94nm.}
\end{figure}

The slope of the torque vs. supercoiling curves at low values of
$\sigma$ (see Fig.\ref{fig:hats}(b)) yields $C_s$, i.e. the effective torsional stiffness of DNA. As shown in Fig.\ref{fig:ccomp} $C_s$ decreases with the force acting on
the molecule which may explain the low values of torsional stiffness initially reported in bulk measurements\cite{selv92}. This was anticipated by Moroz and Nelson\cite{moroz_torsional_1997} who ascribed this variation to a renormalisation of the bare torsional stiffness $C$ by torsional fluctuations (that become more important at low forces). Their estimate of this effect ($ C_s = C \left[ 1-({C}/{4A})\sqrt{{k_B T}/{A F}} \right] $ obtained by a perturbation expansion at high forces) is shown in Fig.\ref{fig:ccomp}. It allows to deduce a value of $C=94 \pm 7$nm, in agreement with previous estimates\cite{stri99_C,bryant_structural_2003,forth_abrupt_2008}.

\begin{figure}[!htb]
\includegraphics[width=9cm]{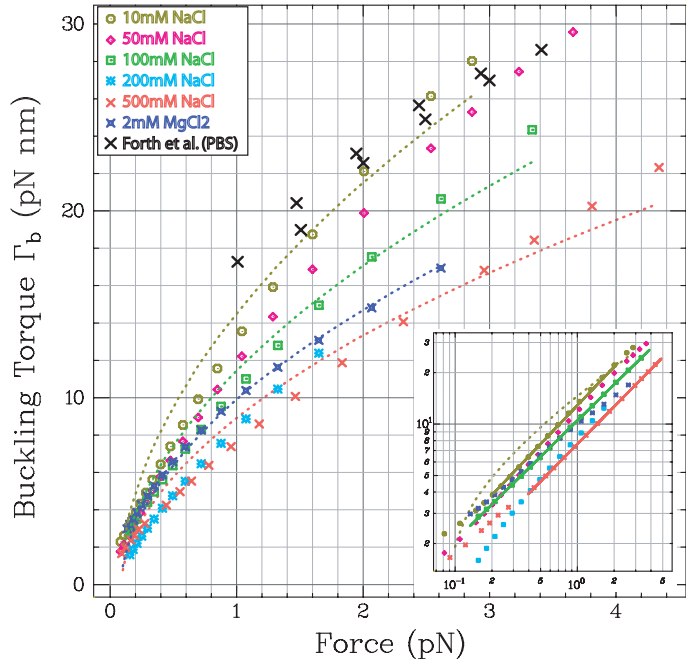}
\caption{\label{fig:bucklingtorque} Buckling torque of DNA for different salt conditions (10mM, 50mM, 100mM, 200mM, 500mM $NaCl$, 2mM
  $MgCl_2$).  Inset: Log-Log plot of the data and power-law fits (continuous lines). The average exponent is $0.72 \pm 0.07$. Dotted lines: best fits to a recent model of J.Marko\cite{marko_torque_2007} (see text).}

\end{figure}

While the value of the DNA bare torsional modulus $C$ does not seem to
vary much with salt, the buckling torque of DNA $\Gamma_b$ appears to
be much affected by the ionic concentration, increasing by as much as a factor 2 at low salt concentrations, see Fig.\ref{fig:bucklingtorque}. The buckling torque increases also with the force\cite{char04} with an approximate power law dependence $\Gamma_b \sim F^{0.72}$. Although the precise values of the buckling torque for DNA may depend slightly on sequence (for example AT tracks are known to form bends that may buckle more easily) one expects the general dependence of $\Gamma_b$ with force and salt to be sequence independent for long enough DNA's. Clauvelin et al. \cite{clauvelin_mechanical_2008} obtain the buckling torque from the experimental slopes of the torque vs. supercoiling curves using an analytically solvable model of plectonemic DNA. The values of buckling torque obtained with their method are very close to the values reported in Fig. \ref{fig:bucklingtorque}.

\begin{figure}[!htb]
\includegraphics[width=9cm]{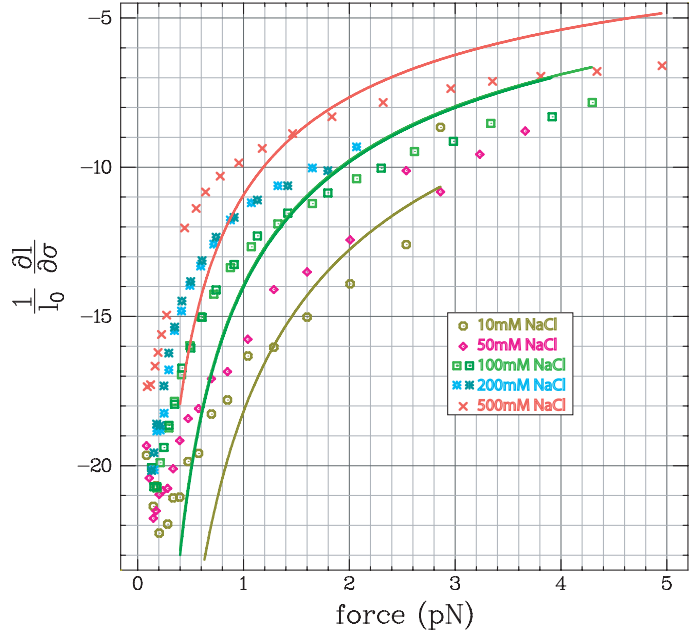}
\caption{\label{fig:slopes}  Slope of the decrease in extension in the
  plectonemic regime (the slope of the BC segment in
  Fig.\ref{fig:hats}(a)) as a function of force and in various salt conditions. The continuous lines are a best fits to a recently proposed model \cite{marko_torque_2007}.}
\end{figure}

J.Marko \cite{marko_torque_2007} has recently suggested a heuristic
model to describe the behavior of a stretched DNA molecule under
twist. In his model, DNA molecules in the plectonemic regime partition
between an unstretched plectonemic supercoil phase with torsional
stiffness $P$ and a stretched and twisted DNA molecule with
persistence length $A$ and effective torsional stiffness
$C_s$. This model is characterized by only three parameters: the DNA
persistence length $A = 50$ nm, its bare torsional stiffness $C
\approx 90$ nm and an unknown plectonemic torsional stiffness $P$, estimated to be
about 26 nm. The model makes a number of predictions on the variation
of extension with $\sigma$ and the variation of $\Gamma_b$ with $F$
that can be compared with experiments. While the predictions of
Marko's model are in qualitative agreement with our observations (see
for example the variation of $\Gamma_b$ with force in Fig.\ref{fig:bucklingtorque}), the model cannot explain all the data
with only three fit parameters (actually only one $P$, since $A$ can
be independently obtained from force-extension measurements, and $C$
can be deduced as explained from Fig.\ref{fig:ccomp}). In particular
as can be seen in Fig.\ref{fig:slopes}, the predictions of the model for the variation of $\partial l / \partial \theta$ as a function of force does not
quantitatively fit the data. In some sense this is not very surprising since a description of the plectonemic phase with
a single force independent torsional stiffness $P$ is an oversimplification that does not take into account for example the
variation of plectonemic radius with force due to entropic repulsion \cite{marko_fluctuations_1994}.

% Specify following sections are appendices. Use \appendix* if there
% only one appendix.
%\appendix
%\section{}

% If you have acknowledgments, this puts in the proper section head.
\begin{acknowledgments}
We would like to thank J. Marko, Z.Bryant and M.Wang for useful discussions and correspondence. This work has been supported by grants from CNRS, the ANR and the EU (BioNano-Switch).
\end{acknowledgments}

% Create the reference section using BibTeX:
\bibliography{torque}

\end{document}